\documentclass[journal,comsoc]{IEEEtran}
\usepackage[T1]{fontenc}
\usepackage{graphicx}

\usepackage{cite}

\ifCLASSINFOpdf
\else
\fi
\usepackage{amsmath}
\usepackage[cmintegrals]{newtxmath}
\begin{document}
%
\title{GEMRank: Global Entity Embedding For Collaborative Filtering}
%
%
%

\author{Arash~Khoeini, 
        Bita~Shams,
        Saman~Haratizadeh,
}

%
%

\markboth{Journal of \LaTeX\ Class Files,~Vol.~14, No.~8, August~2015}%
{Shell \MakeLowercase{\textit{et al.}}: Bare Demo of IEEEtran.cls for IEEE Communications Society Journals}
%



\maketitle

\begin{abstract}
Recently, word embedding algorithms have been applied to map the entities of recommender systems, such as users and items, to new feature spaces using textual element-context relations among them. Unlike many other domains, this approach has not achieved a desired performance in collaborative filtering problems, probably due to unavailability of appropriate textual data. In this paper we propose a new recommendation framework, called GEMRank that can be applied when the user-item matrix is the sole available souce of information. It uses the concept of profile co-occurrence for defining relations among entities and applies a factorization method for embedding the users and items. GEMRank then feeds the extracted representations to a neural network model to predict user-item like/dislike relations which the final recommendations are made based on. We evaluated GEMRank in an extensive set of experiments against state of the art recommendation methods. The results show that GEMRank significantly outperforms the baseline algorithms in a variety of data sets with different degrees of density.
\end{abstract}


%
\IEEEpeerreviewmaketitle

\section{Introduction}

Recommendation Systems help users to find relevant items based on their preferences. Many prominent recommendation systems are using Collaborative Filtering (CF) for making recommendations ( \cite{Goldberg:1992:UCF:138859.138867}). In Collaborative Filtering approach, previous user-item interactions are analyzed to estimate a target user's unknown ratings or preferences. The idea behind CF simply is that users with similar tastes in the past will choose similar items in the future.

Matrix Factorization (MF) is a traditional method of recommendation that has been used massively in Collaborative Filtering (\cite{Koren:2009:MFT:1608565.1608614}). In this method, the user-item rate matrix is factorized to map the items and users to a new latent feature space.  The factorization is done so that the dot product of the latent vectors of a user $u$ and an item $i$ is equal to the rate that $u$ has given to $i$. In other words, the representation mapping practically is not completely separated from the rate prediction, but instead, the former works as a submodule of the latter. This phenomenon can somewhat limit the prediction power of the model: to achieve more flexibility in prediction, the system may have to represent entities with unreasonably long vectors. Another defficiency of the approach is that it maps users and items to the latent feature space at the same time, while they are different entities with different degrees of complexity. An alternate approach that we suggest to use, is to map users and items separately to new possibly different feature spaces. As we will explain later, this can be done by defining simpler kinds of relations, compared to user-item ratings for example, among entities of the same type.

Recently, some word embedding algorithms, have been used in recommendation area for vectorizing the entities of the system(\cite{ozsoy:2016} \cite{barkan2016item2vec}). Like MF methods, deep learning techniques usually seek to find vector representations for users and items, and use them to make recommendations\cite{zheng2017joint}, however they do it mainly by exploiting available textual and other forms of unstructured data. 

A well known word representation learning method is Word2Vec \cite{mikolov2013distributed} which captures representaions for words of a language based on the neighboring words in sentences in which they appear. These technique is usually based on analyzing the co-occurrence of words in the same context, and trying to map words to new vector representations that somehow reflect the context in which each word usually appears. This context-based strategy for word embedding has been shown to be successful in capturing some semantic relations that exist in the language. Roughly speaking, it is expected that in the new feature space a word is closer to its synonyms than other words. This phenomenon has made the deep learning methods good candidates for semantic-preserving summarization of textual data in many applications including recommender systems: If the entities of the recommendation system, such as users or items, are textually defined, explained or reviewed, then a word embedding method may be quite useful in mapping those entities to vector representations that reflect the true similarities among them successfully. Clearly such representations can be helpful when making simple neighbor based, or more sophisticated model based, recommendations. Especially in such an approach, unlike matrix factorization, the vectorization module does not directly depend on the rate prediction function, and that gives more flexibility to the recommendation system.

Although this approach seems promising for recommendation (\cite{shin2014recommending},\cite{grbovic2015commerce}), it has failed to outperform other state of the art collaborative filtering methods, where there is no textual data available to be used for a user-item embedding task. It seem that the main problem in such a situation is that it is not clear how to map the elements of the system to a new feature space, so that the natural similarities among them are captured in their new vector representations. In collaborative filtering domain, unlike the domain of text analysis, the concepts of words, sentences (that is a certain sequence of words), or a word's context (that is usually defined to be its adjacent words in a sentence) do not exist. Since they are basic elements of a standard word embedding method, such methods can not directly be applied to this domain. For example Word2Vec \cite{mikolov2013distributed} defines a window size and assumes that the context of a certain word are the words that appear in that window before or after it in a sentence. Unfortunately it is not a straightforward task to model a recommendation problem in a way that supports these assumptions. So we need to define appropriate counterparts for those concepts as well as an appropriate mapping procedure, before we can use the idea of context-based word embedding for a similarity preserving mapping of the elements of the system to vector representations.  

In this paper we investigate the concept of proximity of the entities based on the choices that users of the system have made so far, and suggest a context-based entity embedding algorithm that derives vector representations for users and items. The concept of the context of entities is defined based on the profiles in which each entity appears and is modelled in the form of a so called profile co-occurrence (PCO) matrix. Based on this definition of context, a matrix factorization approach is then suggested for extracting basic vector representations for users and items. These representation is then fed to a feed-forward neural network for like/dislike prediction. Finally this neural network will act as an estimator and predicts the probability that a user will like an item that is used for recommendation.
The contributions of this research can be summarized as follows:

\begin{itemize}
\item We introduce a new recommendation algorithm that outperforms the state of the art recommendation methods.
\item We introduce the profile co-occurrence matrix, to represent the relations among entities.
\item We suggest to use a factorization based approach for embedding users and items in different steps, in collaborative filtering problems. In this approach the naturally simpler entities of the system are vectorized first and other more complex entities are defined based on them.  
\end{itemize}
The rest of the paper is organized as follows: first we review the related works in the next section, then we propose a new recommendation framework called GEMRank in section 3.  In section 4 the experiments and results are reported. Some experimental observations are discussed in section 5 and the paper is concluded in section 6.

\section{Related Work}
Two main approaches in collaborative ranking are Neighbor-based Collaborative Ranking (NCR) and Matrix Factorization Collaborative Ranking (MFCR). 
One of the first NCR algorithm was EigenRank \cite{liu2008eigenrank} which infers a total ranking based on pairwise preferences of users similar to the target user. EigenRank makes recommendations in two steps: It first computes user similarities using Kendall correlation, based on the agreements and disagreements of users over their preferences. Then it estimates a preference matrix whose elements are a weighted linear combination of neighbors' preferences. Finally, EigenRank uses a greedy or a Markov-based approach to infer a total ranking over items for each user. The same approach has been adopted by many other NCR algorithms with slight modifications [\cite{yang2009cares},\cite{meng2011wsrank}]. For example VSRank \cite{wang2014vsrank} improves Kendall similarity measure by considering importance of each pairwise comparison in similarity calculation or NN-GK  \cite{kalloori2016pairwise} combines a different ranking method to Kendall similarity measure to find rankings.

Traditional NCR algorithms usually suffer from sparsity of the dataset as the similarity calculation can be hard when there is not enough information about users available. Graph based methods try to solve this problem by modeling the data as a graph in order to estimate the distances more accurately when the data is sparse. These methods first construct a graph to represent data and then make recommendations by analyzing the graph. In \cite{yao2013personalized} different types of nodes and a multi-layer structure have been used to make context-aware recommendation through a random walk in the graph. SibRank  \cite{shams2016sibrank} uses a signed bipartite preference network for representing the data and analyzes it using a signed version of Personalized PageRank to capture users' similarities. Among more recent approaches, GRank \cite{shams2017graph} is an state of the art method which uses personalized PageRank over a tripartite preference network to directly infer the total ranking of items. GRank may use unreliable paths that are inconsistent with the general idea of similarity in neighborhood collaborative ranking. ReGRank \cite{shams2018reliable} ranks items based on reliable recommendation paths that are in harmony with the semantics behind different approaches in neighborhood collaborative ranking.

Matrix Factorization Collaborative Ranking tries to find hidden factors based on which users rank items. CofiRank \cite{weimer2008cofi} was the first algorithm which used matrix factorization to optimize a rank-oriented metric. It extracts latent representations in order to optimize a structured loss function. ListRank \cite{shi2010list} tries to learn the latent factors to estimate the top-1 probabilities of items in order to improve the ranking quality. BoostMF \cite{chowdhury2015boostmf} aggregates boosting and matrix factorization algorithms to learn the latent factors which optimize the Top-N recommendation.  PushAtTop \cite{christakopoulou2015collaborative} is another Matrix Factorization algorithm which weights pairwise comparisons according to their position in the total ranking of items for users. In \cite{wei2017collaborative} a deep model is trained to solve the cold start problem for users with zero or few rating records. It first uses a specific deep neural network SADE to extract the content-based features for items. Then it uses a modified version of timeSVD++ to use the content features for prediction of ratings for cold start items.
 
Representation learning and deep learning methods are among the novel approaches used in recommendation systems. Basically, word embedding techniques learn a low-dimensional vector space representation of words in an unsupervised manner. These approaches are recently gaining more attention, since they have shown good performance in a broad range of natural language processing (NLP) scenarios.

Many attempts have been made to use word embedding methods on textual data to generate user/item representations (\cite{musto2015word}, \cite{zheng2017joint}). Since the textual information, like reviews, are not always available, some researchers have tried to use word embedding methods based on other kinds of raw data. For example, \cite{barkan2016item2vec} and \cite{ozsoy:2016} are among the first recommendation methods that adopt word embedding in non-textual datasets. Ozsoy \cite{ozsoy:2016} considers past check-ins of users as words while each sentence is constructed as a sequence of all the venues a user has visited. Then these sentences are fed to Skip Gram \cite{mikolov2013distributed} to make dense vector representations for each venues and finally to make recommendations using neighborhood methods. In \cite{grbovic2015commerce} the name of the users' purchased items are used as the words to form a sentence and each item is mapped to a new feature space. The similarities among the resulting vector representations of items are then used for recommendation. Unfortunately none of these attempts led to any state of the art results due to the fact that the data they have used does not naturally support the properties of textual data, like the natural order or the proximity of the words in a sentence, that form the word-context relations in the textual data.  Modeling such a data as text, can mislead the embedding process and result in inappropriate vector representations that cause poor recommendation performance in later steps.

\section{GEMRank}
In this section we introduce GEMRank, that is a framework for embedding users and items in a collaborative filtering task and use their vector representations for like/dislike prediction. Usually matrix factorization methods in recommendation systems factorize user-item matrix to capture latent factors for users and items. Those representations are then directly used for predicting the unknown user-item rates. GEMRank, on the other hand, separates the concept of element embedding from the rate prediction. It tries to learn representations for the elements of the system based on an element-context relation. Then it uses the resulting vector representations for training a preference prediction model. So, GEMRank has two main phases: In the first phase it generates an item-item or user-user  relation matrix and factorizes it to map the entities to their vector representations. The elements  that are vecotrized in this step are called the basic element and the user-item matrix is used to derive vector representations for the other elements of the system based on the representations of the basic elements. 

In the second phase, GEMRank feeds the generated embeddings to a feed forward neural network and trains it to predict user-item like/dislike relations. This neural network is later used for predicting the unknown like/dislike relations among users and items, based on which the final recommendation list is derived for each target user.

The main concept using which GEMRank defines the element-context relations and generates the vector representations for the entities is the profile co-occurrence concept. The profile of each user in a collaborative filtering task contains the items that user has rated/liked/.etc. Similarly the profile of an item contains the users that have rated, bought or liked that item. GEMRank interprets this information as a clue for a kind of relation among those elements: When two items appear in a user profile together, they are probably liked by that user, or at least, that user has expected them to suit his taste for some reason. GEMRank uses this kind of relations, as a counterpart to the word-context relations in word embedding approaches. In other words, the number of times that two elements appear in a profile together can somehow reflect the proximity of those elements. GEMRank interprets the other elements that occur in a profile together with a certain element as that element's context. The contexts in which items (users) appear can be aggregated to reflect the natural associations among items (users). 
GEMRank takes one of two kinds of entities, that are users and items, as the basic entity of the system. It then uses the element-context relations for extracting vector representations for the basic entity and then calculates the representations for the other entities based on the embeddings of the basic entity. Although either users or items can be taken as the basic entity, it seems normal to pick items, as they are naturally simpler entities compared to users. So, from here on, we will assume the items as the basic entity and the user profiles as their contexts. However the procedure is the same when assuming users as the basic entity and item profiles as their contexts. 
It may be intriguing to directly interpret the items as words and the users' profiles as sentences and feed them to a text based embedding system. However as we mentioned before, the items in the user profiles does not appear in a certain order, or if they do, that order does not reflect the concept of sequence as exists for the words of a sentence.  So we need to represent the contexts of items in a way that supports the concept of co-occurrence without assuming any sequence among them. 

GEMRank defines that items i  and j are in the context of each other if they have appeared together in a user's profile. Then it constructs the item profile co-occurrence (PCO) matrix $P$, in which the entry $P_{i,j}$ is the number of times that item $i$ has appeared in the context of an item $j$. So $P_{i,j}$ is equal to number of users who have both items $i$ and $j$ in their profiles (equivalently, the user co-occurrence profile can be defined when the basic entity is user rather than item) .  For example in a movie recommender system, it can be the number of users who have watched both movies $i$ and $j$. 

GEMRank then assumes that the occurrence of an item in another items context depends on some unknown features of those items from the view point of the users. So it tries to find vector representations for items based on those hidden features by factorization the PCO matrix P.

After composing matrix P, we need to use it as a clue to capture item representations. We can do it by simply factorizing matrix P into two matrices: target-item matrix and contect-item matrix. If we consider the vector representation of a target-item $i$ as $v_i$ and the vector representation of a context-item $j$ as $v'_j$, then GEMRank tries to find $v_i$ and $v'_j$ in a way that equation below holds:

\begin{equation} \label{objective}
    v_i^T v'_j  = f(P_{ij}) ,
\end{equation}
where function f is an smoothing function which can be defined as equation \ref{func_f}

\begin{equation} \label{func_f}
   f(x)  = \left\{ \begin{array}{rcl} 
        log(P_{ij})  & \mbox{for} & z>1 \\
        P_{ij}      & \mbox{for} & z \le 1
        \end{array}\right.
\end{equation}
note that f acts as a smoothing function on P. 

To capture all $v$ and $v'$ vectors, the following cost function is minimized:

\begin{equation} \label{cost}
    J = \sum ^ I _ {i,j=1} (v_i^T v'_j - f(P_{ij}))^2 ,
\end{equation}
where $I$ is the number of all items. 

By minimizing the equation \ref{cost}, GEMRank finds two representations for each item: one for it as a target-item and one for it as a context-item. Although any of these two vectors can be used as that item's vector representations, our experiments showed that using just one vector or the sum of the two vectors does not yield significant change in results. 

When the item vectors are found, GEMRank constructs the vector representations for users based on the intuition that each user's taste can be defined as the aggregation of all item's that he has in his profile. So the vector representation for each user $i$ is built as follows: 

\begin{equation} \label{uservector}
    u_i = \sum_{k \in M_i} v_k ,
\end{equation}
Where $u_i$ is the vector representation for user $i$ and $M_i$ is list of all the movies that user i has watched. 

When the recommendation is based on binary implicit feedback data, the above definition for the users' vector representation can be applied. For situations in which the data contains information like grades of feedback or rates, the vector representation of a user is defined as follows:

\begin{equation} \label{uservector2}
    u_i = \sum_{k \in M_i} (r_{i,k} - \lambda_{i}) v_k ,
\end{equation}
where  $r_{i,k}$ is the rate user i has given to movie k and $\lambda_{i}$ is the average of user i's ratings.

Once vector representations for users and items are generated, a neighborhood recommendation algorithm can use them to find similar users or items and recommend items to users based on those similarities. Here we can use the same approach, but we will show that by training a feed forward neural network for like/dislike prediction  we can do much better in inferring the possible interest of each user to each item.

\begin{figure*}[ht]
    \caption{MLP architecture}
    \label{mlp}
      \includegraphics[width=\textwidth]{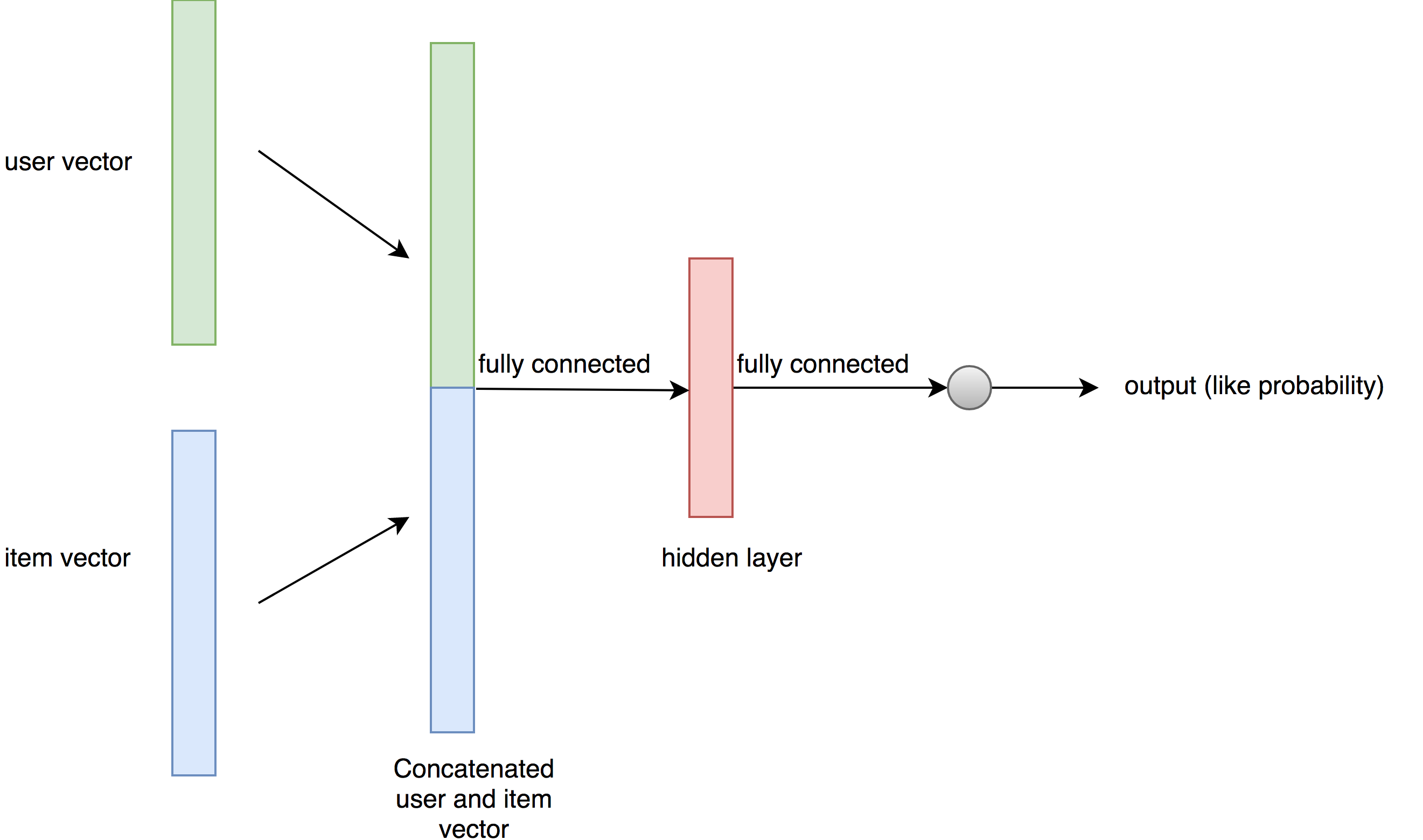}
\end{figure*}

So in the second phase the algorithm, GEMRank uses a feed forward neural network with one hidden layer. This  neural network is a Multi Layer Perceptron (MLP) with on hidden layer which learns to predict the interest for each (item, user) pair. 
As we can see in figure \ref{mlp}, the input layer of this neural network is concatenated vectors of user and item. In the next layer, which is the hidden layer of our neural network, these two vectors will interact with eachother in this hidden layer and produce final result will be the probability that a user will like an item. If we consider this probability as $\Phi$, this neural network's goal is to estimate the following funciton

\begin{equation}
  f( \Phi | v_j,u_i,\Theta),
\end{equation}
where $\Theta$ is parameters of neural network.

To achieve this goal we train the neural network with concatenated vectors of the user and item as input and the interest level of user to that item as output. We consider interest level as the user rate divided by maximum possible rate. For example if a user has rated and item 4 out of 5, her interest level is considered $\frac{4}{5}$. The output neuron uses a sigmoid activation function that gives GEMRank the opportunity to rank items for each user based on the levels of activations observed in the output. The final function of this neural network will be as below:

\begin{equation}
  \Phi  = W' ReLU(W \left[ \begin{array}{c}u_i \\ v_j \end{array} \right] + b) + b'.
\end{equation}

\section{Experiments}
GEMRank is experimented and tested on MovieLens rating data. We used the rating data of these datasets and ignored timestamps. MovieLens-100K contains 100 thousands rating data of 943 users with at least 20 rated movies for each user and there are 1682 distinct movies in it. MoveiLens-1M contains 1 Million ratings by 6040 users with at least 20 rated movies for each user. There are 3952 distinct movies in MoveiLens-1M.

For the first phase, we tried vector with 30 and 120 dimensions while the best results were achieved at 100. For our neural network we chose 25 neurons for both hidden layers and 10 neurons for the shared layer. The activation functions in hidden and shared neurons are ReLu while, as we mentioned, sigmoid activation function has been used for the output neuron.

Different sizes of user profiles are used to compare GEMRank with baseline methods. For each user, profile is the movies she has rated and a user's profile size is the number of items in his train set. For each user, a fixed number (UPL) of ratings are selected from user profiles for training and the remaining ratings are left for the test process. We used UPL size of 10, 20 and 50 in our experiments.

\subsection{Baseline Algorithms}\
GEMRank is compared with both model-based and neighbor-based algorithms:
\begin{itemize}
     \item SibRank \cite{shams2016sibrank} that uses a signed bipartite preference network to capture users similarities and resolves the sparsity issue of NCR algorithms. 

    \item GRank \cite{shams2017graph}, which uses personalized PageRank over a tripartite preference network to rank items
    
    \item ListRank \cite{shi2010list}  which is a state-of-the-art ranking-oriented Collaborative Filtering approach.

    \item CofiRank \cite{weimer2008cofi}, that uses a Matrix Factorization Collaborative Ranking approach.

    \item PushAtTop \cite{christakopoulou2015collaborative} which is a Matrix Factorization algorithm which weights pairwise comparisons according to their position in the total ranking of items for users. PushAtTop contains three different algorithms: push-inf, push-p, push-reverse.

\end{itemize}

\subsection{Parameter Setting}
In GEMRank, we set the size of vector dimensions to 100, so the input layer of neural network will have 200 neurons. We used 5 percent of the training data as validation set for choosing the best size ( 5, 10, 15, 20 or 25) for hidden layer  of the neural network. The activation function of the  neourons of the hodden layer is ReLU. We also used a Dropout of 0.5 to avoid overfiting in smaller UPL sizes.

Parameters of the baseline algorithms are set as follows:

\begin{itemize}

    \item In SibRank, neighborhood size of 100 and $\alpha = 0.85$  are used. These parameter are reported as the best parameters in the original paper \cite{shams2016sibrank} .

    \item In GRank $\alpha = 0.85$ based on the original paper \cite{shams2017graph}.

    \item In ListRank, dimensionality of latent features (d) is 10 and regularization parameter is set to 0.1.

    \item In CofiRank, publicly available framework is used and the best parameters reported from \cite{weimer2008cofi} is used.

    \item For three different algorithms of PushAtTop, All the optimal algorithms reported in \cite{christakopoulou2015collaborative} are
    used.
\end{itemize}

\subsection{Evaluation Metric}
A metric for measuring the performance of rank-based recommender systems is NDCG, which is used to evaluate top-n recommendations.
More top-n recommendations are similar to real top-n items ranked by user, the NDCG is closer to 1. NDCG for top-n items are calculated 
as below:

\begin{equation} \label{NDCG}
    NDCG_u = \frac{1}{\beta_u} \Sigma^{top-n}_{i=1} \frac{2^{r^u_i} - 1} {log^i_2 +1}
\end{equation},

where $r^u_i$ is the rate of the $i^{th}$ ranked item and $\beta$ is a normalizing factor that makes sure that NDCG is between 0 and 1.

\subsection{Result} 
We implemented two versions of GEMRank. In the first version, we factorize the item PCO matrix as explained above and we call it item-based GEMRank. In the second version, the user PCO matrix is factorized and then an item's representations is inferred using representations of users who have watched it.
\begin{table*}[ht] \caption{Performance comparison in terms of NDCG between GEMRank and state of art recommendation algorithms on MovieLens 100k dataset.}
    \label{result100k}
    \centering
    \resizebox{\textwidth}{!}{

\begin{tabular}{ |c||c|c|c|c|c|c|  }

    \hline
     &              \multicolumn{2}{c}{UPL = 10}             &      \multicolumn{2}{c}{UPL = 20}       & \multicolumn{2}{c|}{UPL = 50} \\
    \hline
    algorithm  & NDCG@5             & NDCG@10             & NDCG@5             & NDCG@10             & NDCG@5             & NDCG@10  \\
    \hline
    GEMRank (item-based)     & \textbf{0.676 $\pm$ 0.008}& \textbf{0.692 $\pm$ 0.006}& \textbf{0.695 $\pm$ 0.009}& \textbf{0.699 $\pm$ 0.009}& 0.713 $\pm$ 0.004& 0.715 $\pm$ 0.003\\
    GEMRank (user-based) &  0.624$\pm$0.010 & 0.654$\pm$0.007 & 0.615$\pm$0.009 & 0.642 $\pm$ 0.004    & 0.602 $\pm$ 0.010  & 0.629 $\pm$ 0.008  \\
    SibRank    & 0.622 $\pm$ 0.010  & 0.650 $\pm$ 0.009   & 0.660 $\pm$ 0.015  & 0.672 $\pm$ 0.012    & 0.711 $\pm$ 0.006  & 0.710 $\pm$ 0.004  \\
    GRank      & 0.595 $\pm$ 0.030  & 0.632 $\pm$ 0.024   & 0.642 $\pm$ 0.022  & 0.657 $\pm$ 0.020    & \textbf{0.719 $\pm$ 0.004}  & \textbf{0.717 $\pm$ 0.005}  \\
    ListRank   & 0.672 $\pm$ 0.005  & 0.693 $\pm$ 0.005   & 0.682 $\pm$ 0.006  & 0.691 $\pm$ 0.004    & 0.687 $\pm$ 0.008  & 0.684 $\pm$ 0.004  \\
    CofiRank   & 0.602 $\pm$ 0.020  & 0.631 $\pm$ 0.013   & 0.603 $\pm$ 0.014  & 0.620 $\pm$ 0.024    & 0.609 $\pm$ 0.010  & 0.616 $\pm$ 0.004  \\
    push-inf   & 0.611 $\pm$ 0.030  & 0.640 $\pm$ 0.023   & 0.629 $\pm$ 0.015  & 0.647 $\pm$ 0.012    & 0.658 $\pm$ 0.003  & 0.667 $\pm$ 0.002  \\
    push-reverse& 0.594 $\pm$ 0.043  & 0.623 $\pm$ 0.033  & 0.621 $\pm$ 0.018  & 0.640 $\pm$ 0.016    & 0.664 $\pm$ 0.012  & 0.668 $\pm$ 0.013  \\
    push-p     & 0.519 $\pm$ 0.021  & 0.561 $\pm$ 0.015   & 0.580 $\pm$ 0.034  & 0.602 $\pm$ 0.029    & 0.681 $\pm$ 0.012  & 0.679 $\pm$ 0.011  \\
    
    \hline
   \end{tabular}
    }
\end{table*}

\begin{table*}[ht] \caption{Performance comparison in terms of NDCG between GEMRank and state of art recommendation algorithms on MovieLens 1M dataset.}
    \label{result1M}
    \centering
    \resizebox{\textwidth}{!}{

\begin{tabular}{ |c||c|c|c|c|c|c| }

    \hline
     &             \multicolumn{2}{c}{UPL = 10}             &     \multicolumn{2}{c}{UPL = 20}        &      \multicolumn{2}{c|}{UPL = 50}     \\
    \hline
    algorithm     & NDCG@5             & NDCG@10            & NDCG@5            & NDCG@10             & NDCG@5            & NDCG@10            \\
    \hline
    GEMRank (item-based) & \textbf{0.726 $\pm$ 0.004} & \textbf{0.730 $\pm$ 0.002} & \textbf{0.725$\pm$ 0.003} & \textbf{0.730$\pm$ 0.003} & {0.755$\pm$ 0.001} & {0.753 $\pm$ 0.0007}   \\
    GEMRank (user-based) & 0.708 $\pm$ 0.002 & 0.717 $\pm$ 0.0002 & 0.706$\pm$ 0.002 & 0.715$\pm$ 0.009 & 0.705 $\pm$ 0.005 & 0.712$\pm$ 0.005  \\
    SibRank       & 0.670 $\pm$ 0.009          &      0.685 $\pm$ 0.001        &      0.701 $\pm$ 0.007    &     0.701 $\pm$ 0.006    &     0.727 $\pm$ 0.016     &     0.723 $\pm$ 0.013  \\
    GRank         & 0.638 $\pm$ 0.022          &      0.654 $\pm$ 0.019        &      0.692 $\pm$ 0.013    &     0.694 $\pm$ 0.014    &\textbf{0.757 $\pm$ 0.019}     
&\textbf{0.755 $\pm$ 0.014}  \\
    ListRank & 0.647 $\pm$ 0.002 &  0.654 $\pm$ 0.002 & 0.683 $\pm$ 0.003 & 0.688 $\pm$ 0.003 &  0.751 $\pm$ 0.002 &  0.751 $\pm$ 0.002  \\
    CofiRank & 0.685 $\pm$ 0.002 &  0.684 $\pm$ 0.001 & 0.676 $\pm$ 0.008 & 0.685 $\pm$ 0.024 &  0.641 $\pm$ 0.005 &  0.644 $\pm$ 0.003  \\
    push-inf      & 0.690 $\pm$ 0.008          &      0.699 $\pm$ 0.002        &      0.691 $\pm$ 0.017    &     0.697 $\pm$ 0.012    &     0.695 $\pm$ 0.011     &     0.695 $\pm$ 0.008  \\
    push-reverse  & 0.640 $\pm$ 0.004          & 0.713 $\pm$ 0.023        &      0.717 $\pm$ 0.029    &     0.713 $\pm$ 0.023    &     0.738 $\pm$ 0.010     &     0.734 $\pm$ 0.010  \\
    push-p        & 0.696 $\pm$ 0.008          &      0.706 $\pm$ 0.003        &      0.699 $\pm$ 0.026    &     0.705 $\pm$ 0.021    &     0.714 $\pm$ 0.025     &     0.714 $\pm$ 0.022  \\
    
    \hline
   \end{tabular}
    }
\end{table*}

Table \ref{result100k} and \ref{result1M} show the result of our experiments and comparisons with baseline algorithms on MovieLens 100k data and MovieLens 1M receptively. As we can see,  GEMRank performs better than baseline algorithms in almost every settings in both data sets.

To observe the impact of our parallel feed forward neural network, we removed the neural network from GEMRank and used a neighbor-based approach that recommends k nearest items to each user. Table \ref{result100_kni} shows the difference in NCDG results when we use parallel feed forward to make recommendations and when we do not use it. The results are presented in table \ref{result100_kni}

\begin{table*}[ht] \caption{Performance comparison in terms of NDCG between GEMRank and GEMRank without MLP phase on MovieLens 100k dataset.}
  \label{result100_kni}
    \centering
    \resizebox{\textwidth}{!}{

\begin{tabular}{ |c||c|c|c|c|c|c| }

    \hline
     &             \multicolumn{2}{c}{UPL = 10}             &     \multicolumn{2}{c}{UPL = 20}        &      \multicolumn{2}{c|}{UPL = 50}     \\
    \hline
    algorithm     & NDCG@5             & NDCG@10            & NDCG@5            & NDCG@10             & NDCG@5            & NDCG@10            \\
    \hline
    GEMRank (item-based using MLP)     & \textbf{0.676 $\pm$ 0.008}& \textbf{0.692 $\pm$ 0.006}& \textbf{0.695 $\pm$ 0.009}& \textbf{0.699 $\pm$ 0.009}& \textbf{0.713 $\pm$ 0.004}& \textbf{0.715 $\pm$ 0.003}\\
    GEMRank (item-based using Simple Approach)  & 0.511 $\pm$ 0.008 & 0.547 $\pm$ 0.005   &      0.540 $\pm$ 0.008    &       0.564 $\pm$ 0.007   &      0.598 $\pm$ 0.001    &    0.600 $\pm$ 0.002 \\
 
    \hline
   \end{tabular}
    }
\end{table*}

\begin{table*}[ht] \caption{Performance comparison in terms of NDCG between GEMRank and user-item GEMRank on MovieLens 100k dataset.}
  \label{result100_user_item}
    \centering
    \resizebox{\textwidth}{!}{

\begin{tabular}{ |c||c|c|c|c|c|c| }

    \hline
     &             \multicolumn{2}{c}{UPL = 10}             &     \multicolumn{2}{c}{UPL = 20}        &      \multicolumn{2}{c|}{UPL = 50}     \\
    \hline
    algorithm     & NDCG@5             & NDCG@10            & NDCG@5            & NDCG@10             & NDCG@5            & NDCG@10            \\
    \hline
    GEMRank (item-based)     & \textbf{0.676 $\pm$ 0.008}& \textbf{0.692 $\pm$ 0.006}& \textbf{0.695 $\pm$ 0.009}& \textbf{0.699 $\pm$ 0.009}& \textbf{0.713 $\pm$ 0.004}& \textbf{0.715 $\pm$ 0.003}\\
    GEMRank (user-item)  & 0.673 $\pm$ 0.010 & 0.690 $\pm$ 0.008   &  0.681 $\pm$ 0.004    &    0.693 $\pm$ 0.001   &      0.685 $\pm$ 0.009    &    0.690 $\pm$ 0.007 \\
    \hline
   \end{tabular}
    }
\end{table*}

\section{Discussion}
As it can be seen in the results item-based GEMRank performs better than matrix factorization methods, CofiRank and PushAtTop, in all experiments. One possible explanation for this is that unlike traditional MF methods that factorize user-item matirices, item-based GEMRank factorizes a matrix that represents relations among items. Since items are simpler entities compared to users, one can expect that this approach will generate more stable and meaningful embeddings that lead to better recommendations.  
Another evidence that supports this explanation, is the superiority of the performance of item-based GEMRank compared to user-based GEMRank in both datasets and in all UPL sizes, as summarized in tables \ref{result100k} and \ref{result1M}. So in general, it seems advisable to first extract the vector representations for items, as the basic entities, and then calculate vector representations for users based on resulted item vectors. 

It is also important to remember that GEMRank is able to use a more sophisticated prediction module, compared to MF methods, for predicting the unknown relations among users and items based on their new vector representations. In table \ref{result100_kni} we can see the difference in the performance of the item based GEMRank when we use neural networks compared to when we use a simple approach instead; where nearest items to users are recommended to them. The results show that using a neural network to learn a function to predict the like/dislike relations among captured representaions massively improves the quality of the final recommendation. Table \ref{result100_user_item} also shows the NDCG performance when we factorize user-item matrix to capture representaions. As we can see, factorizing item-item matrix performs better specially in more dense datasets (higher UPL).

As we can see two main competitors of GEMRank are SibRank and GRank which both are neighbor based methods. In sparser datasets (smaller UPL values) the neighbor based methods face the problem of insufficient local information and GEMRank significantly outperforms them, thanks to its generalization power. The more interesting observation is that although SinRank and GRank outperform other model-based methods in dense data sets (bigger ULP sizes) with a large margin, GEMRank can manage to perform quite closely to them in those data sets too. In other words, despite the face that traditional MF methods fail to capture the local information that is available in dense datasets GEMRank can efficiently use that information for improving its recommendation power. The reason is probably that it has the required flexibility to capture the local patterns that are usually used by neighbor-based methods for making better recommendations in information-rich environments.

  %

\section{Conclusion}
In this paper we proposed a novel approach which, unlike other traditional matrix factorization methods, infers vector representations for users and items based on some element-context relation among them and then uses the resulting vectors in another module for training a prediction model. The entity embedding is done by factorizoing a so-called co-occurrence matrix that reflects the element-context relations among the basic entities of the system. The vector representations are then fed to a couple of feed forward neural networks that jointly predict the like/dislike relations among users and items.
The co-occurrence matrix that is factorized by GEMRank reflects an element-context relation that is a simpler relation compared to the user-item relation that is the base of traditional matrix factorization methods. Also, using items as the basic entities in the embedding process led to better results, probably due to simpler nature of the items compared to users. The other advantage of the GEMRank approach is that it separates the embedding module from the prediction module. So the vectors representing the entities can be kept simpler while the system can use a sophisticated prediction model to capture complicated user-item relations. According to our experiments, using this approach has given GEMRank the generalization power of model based methods in sparse data sets as well as the flexibility of neighbor-based methods in dense data sets.
For future studies we suggest to use extra information sources such as content, to extract better representations for entities. Also  trying to improve the performance of the prediction module, possibly by using deep neural netowrks, seems to be a promising direction for research.

\ifCLASSOPTIONcaptionsoff
  \newpage
\fi

\bibliographystyle{IEEEtran}
\bibliography{bare_jrnl_comsoc}

\begin{IEEEbiography}[{\includegraphics[width=1in,height=1.25in,clip,keepaspectratio]{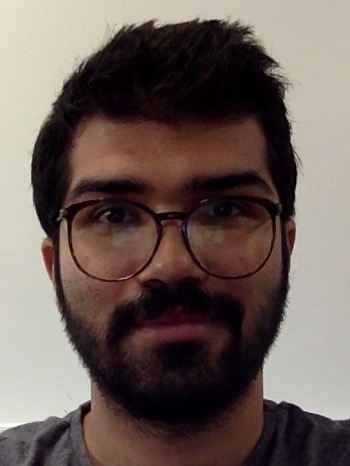}}]{Arash Khoeini}
got his bachelor of Computer Engineering at Kharazmi University in 2015.
Currently, He is a M.Sc. student of Computer Science (Decision Science and Knowledge Engineering) at University of Tehran, Iran. During his master, As a member of Knowledge Discovery and Data Mining lab, Arash is working on Recommendation Systems and Deep Learning. His M.Sc. thesis focus on user and item representations in Recommendation Systems and he is developing novel approaches to improve user/item representation. His research interests includes Deep Learning, Recommendation Systems and Data Mining.
\end{IEEEbiography}

\begin{IEEEbiography}[{\includegraphics[width=1in,height=1.25in,clip,keepaspectratio]{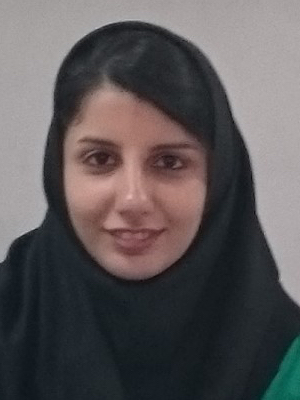}}]{Bita Shams}
received her PhD degree in Information Technology at University of Tehran in 2018. Before her Ph.D., She got her M.Sc. from University of Tehran in 2011, in medical Information Technology.
Currently, the main core of her researches are recommender system and graph mining. Her Ph.D. thesis was entitled as "A Network Oriented Approach To Neighbor-based Collaborative Ranking" in which she designed and exploited novel graph structures that represent users' priorities and pairwise comparisons. She is also interested in other aspects of recommender systems such as shilling attack detection, preference elicitation, trust-ware and serendipitous recommendation.
\end{IEEEbiography}

\begin{IEEEbiography}[{\includegraphics[width=1in,height=1.25in,clip,keepaspectratio]{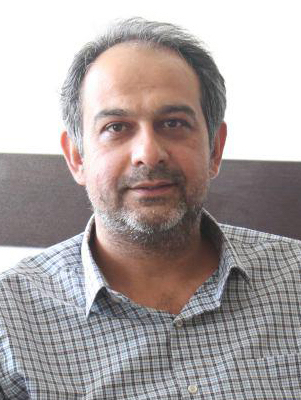}}]{Saman Haratizadeh}
received his Ph.D. degree in computer engineering from Sharif university of technology, Tehran, Iran, in 2007. He is currently an assistant professor and leads the Knowledge Discovery and Data Mining laboratory at the University of Tehran, Iran. His research interests include machine learning, network science, recommender systems, and financial data analysis.
\end{IEEEbiography}

\end{document}